\documentclass[prl,preprint,amsmath,amssymb]{revtex4}
\usepackage{graphicx}
\usepackage[justification=centering]{caption}

\begin{document}

\title{Monogamy relations of all quantum correlation measures for multipartite quantum systems}

\author{Zhi-Xiang Jin$^{1}$}
\author{Shao-Ming Fei$^{1,2}$}

\affiliation{$^1$School of Mathematical Sciences, Capital Normal University,
Beijing 100048, China\\
$^2$Max-Planck-Institute for Mathematics in the Sciences, 04103 Leipzig, Germany}

\bigskip

\begin{abstract}
The monogamy relations of quantum correlation restrict the sharability of quantum correlations in multipartite quantum states. 
We show that all measures of quantum correlations satisfy some kind of monogamy relations for arbitrary multipartite quantum states.
Moreover, by introducing residual quantum correlations, we present tighter monogamy inequalities that are better than all the existing ones.
In particular, for multi-qubit pure states, we also establish new monogamous relations based on the concurrence and concurrence of assistance
under the partition of the first two qubits and the remaining ones.
\end{abstract}

\maketitle

Quantum correlation between many parties is an important quantum phenomena, which plays significant roles in quantum information processing \cite{CS,CGC,JZC,FFF,MAV,LRA}.  Therefore, quantifying quantum correlations becomes more and more important. Recently, some researchers explore the relations shared by more than two parties and expect the measurement of quantum correlation satisfies a monotonic under some local quantum operations \cite{rpmk,kah}. 

The monogamous nature of quantum correlations plays a key role in the security of quantum cryptography \cite{AKE,CBHSB,KKA,SSS,bt,bf,mdr,hqy,csm,am,NGWH}. Monogamy relations are not always satisfied by a correlation measure, for example, the entanglement of formation \cite{CBHSB} which quantifies the amount of entanglement required for preparation of a given bipartite quantum state. However, although the concurrence \cite{concurrence} and entanglement of formation do not satisfy the monogamy inequality $\mathcal{E}_{A|BC}\geq \mathcal{E}_{AB} +\mathcal{E}_{AC}$ ($\mathcal{E}_{A|BC}$ implies the entanglement between $A$ and $BC$), the authors have proved that the $\alpha$th ($\alpha\geq2$) power of concurrence and $\alpha$th ($\alpha\geq\sqrt{2}$) power of entanglement of formation for $N$-qubit states satisfy the monogamy inequalities \cite{ZXN}. In \cite{JF} a tighter monogamy relation for $\alpha$th ($\alpha\geq2$) power of concurrence has been presented.
It has been shown that the information-theoretic quantum correlation measure, quantum discord \cite{HOWH}, can violate the monogamy relations \cite{RPAK,GLGP,RPAKA,XJRHF},
but a monotonically increasing function of the quantum discord could satisfy the monogamy relation for three-qubit pure states \cite{YKBN}.

In this paper, we first show that all quantum correlation measures satisfy some kind of monogamy relations for arbitrary multipartite quantum states.
Then we introduce the residual quantum correlations, and present tighter monogamy inequalities that are better than all the existing ones.
For multi-qubit pure states, we establish new monogamous relations based on the concurrence and concurrence of assistance
under the partition of the first two qubits and the rest ones.

\smallskip

Let $\mathcal{Q}$ be an arbitrary quantum correlation measure of bipartite systems. The quantum correlation measure $\mathcal{Q}$ is said to be monogamous for state $\rho_{AB_1B_2\cdots B_{N-1}}$, if  \cite{ARA},
\begin{eqnarray}\label{}
\mathcal{Q}(\rho_{AB_1})+\mathcal{Q}(\rho_{AB_2})+\cdots+\mathcal{Q}(\rho_{AB_{N-1}})\leq \mathcal{Q}(\rho_{A|B_1B_2\cdots B_{N-1}}),
\end{eqnarray}
where $\rho_{AB_i}$, $i=1,...,N-1$, are the reduced density matrices, $\mathcal{Q}(\rho_{A|B_1B_2\cdots B_{N-1}})$ denotes the quantum correlation $\mathcal{Q}$ of the state $\rho_{AB_1B_2 \cdots B_{N-1}}$ under bipartite partition  $A|B_1B_2\cdots B_{N-1}$. For simplicity, we denote $\mathcal{Q}(\rho_{AB_i})$ by $\mathcal{Q}_{AB_i}$, and $\mathcal{Q}(\rho_{A|B_1B_2 \cdots B_{N-1}})$ by $\mathcal{Q}_{A|B_1B_2\cdots B_{N-1}}$.
One can define the $\mathcal{Q}$-monogamy score for the $N$-partite state $\rho_{AB_1B_2 \cdots B_{N-1}}$,
\begin{eqnarray}\label{ll}
\delta_{\mathcal{Q}}=\mathcal{Q}_{A|B_1B_2 \cdots B_{N-1}}-\sum_{i=1}^{N-1}\mathcal{Q}_{AB_i}.
\end{eqnarray}
 $\delta_{\mathcal{Q}}\geq0$ implies Eq. (\ref{ll}) satisfied the monogamy relation. For example, the square of the concurrence has been proved satisfied the monogamy inequality \cite{AKE,SSS} for all multi-qubit states.
However, other measures such as entanglement of formation, quantum discord are failed to be monogamous for pure three-qubit states \cite{GLGP, RPAK}.

From the results in Ref. \cite{SPAU}, given any quantum correlation measure, one can always find a function of the given measure that is monogamous for the same state.
For arbitrary dimensional tripartite states, it has been shown that there exists a  $\beta_{\min}(\mathcal{Q})\in R$ such that for any $\gamma\geq \beta_{\min}(\mathcal{Q})$,  $\mathcal{Q}$ satisfies 
\begin{eqnarray}\label{aq}
\mathcal{Q}^\gamma_{A|BC}\geq\mathcal{Q}^\gamma_{AB}+\mathcal{Q}^\gamma_{AC}.
\end{eqnarray}

In the following, we denote $\beta=\beta_{\min}(\mathcal{Q})$ the minimal value such that $\mathcal{Q}$ satisfies the above inequality for convenience. Generalizing the conclusion (\ref{aq}) to the $N$ partite case, we have the following result.

{\bf[Theorem 1]}. For any $d\otimes d_1\otimes\cdots \otimes d_{N-1}$ state $\rho_{AB_1B_2\cdots B_{N-1}}$, we have
\begin{eqnarray}\label{th1}
\mathcal{Q}^\alpha_{A|B_1B_2\cdots B_{N-1}}\geq \sum_{i=1}^{N-1}\mathcal{Q}_{AB_i}^\alpha,
\end{eqnarray}
for $\alpha\geq\beta$, $N\geq 3$.

{\sf [Proof]}. The Eq. (\ref{th1}) reduces to Eq. (\ref{aq}) for $N=3$. Suppose the Theorem 1 holds for $N-2$. Then, if we consider the state $\rho_{AB_2\cdots B_{N-1}}$, we have  
$$\mathcal{Q}^\alpha_{A|B_2\cdots B_{N-1}}\geq  \sum_{i=2}^{N-1}\mathcal{Q}^\alpha_{AB_i},$$
for any for $\alpha\geq\beta$.

As follows from applying Eq. (\ref{aq}) for the tripartite subdivision $A|B_1|B_2\cdots B_{N-1}$, we have 
\begin{eqnarray*}\label{pfth11}
\mathcal{Q}^\alpha_{A|B_1B_2\cdots B_{N-1}}&&\geq \mathcal{Q}^\alpha_{AB_1}+\mathcal{Q}^\alpha_{A|B_2\cdots B_{N-1}}\\\nonumber
&&\geq\mathcal{Q}^\alpha_{AB_1}+\sum_{i=2}^{N-1}\mathcal{Q}^\alpha_{AB_i}\\\nonumber
&&=\sum_{i=2}^{N-1}\mathcal{Q}^\alpha_{AB_i},
\end{eqnarray*}
for any $\alpha\geq\beta$.
\hfill \rule{1ex}{1ex}

Theorem 1 gives a general result for arbitrary measure of quantum correlations. However, such relations can be further improved by tightening the lower bound of the inequality (\ref{th1}).
Similar to the three tangle of concurrence, for tripartite quantum states $\rho\in H_A\otimes H_B\otimes H_C$, we define the residual quantum correlation as a function of $\alpha$,
\begin{eqnarray}\label{re}
\mathcal{Q}^{\alpha}_{A|B|C}(\alpha)=\mathcal{Q}^{\alpha}_{A|BC} -\mathcal{Q}^{\alpha}_{AB}-\mathcal{Q}^{\alpha}_{AC},~~~\alpha\geq\beta.
\end{eqnarray}
In the following, we denote $\mathcal{Q}^{\alpha}_{A|B|C}=\mathcal{Q}^{\alpha}_{A|B|C}(\alpha)$ for convenience. Now consider a $d\otimes d\otimes d\otimes d$ state $\rho_{AB_1B_2B_3}$.
Define $\mathcal{Q}^{\alpha}_{A|B^\prime_1|B^\prime_2}=\mathrm{max}\{\mathcal{Q}^{\alpha}_{A|B_{1}|B_{2}}, \mathcal{Q}^{\alpha}_{A|B_{1}|B_{3}}, \mathcal{Q}^{\alpha}_{A|B_{2}|B_{3}}\}$,
where $B^\prime_1$ and $B^\prime_2$ stand for two of $B_1$, $B_2$ and $B_3$ such that
$\mathcal{Q}^{\alpha}_{A|B^\prime_1|B^\prime_2}=\mathrm{max}\{\mathcal{Q}^{\alpha}_{A|B_{1}|B_{2}}, \mathcal{Q}^{\alpha}_{A|B_{1}|B_{3}}, \mathcal{Q}^{\alpha}_{A|B_{2}|B_{3}}\}$.

{\bf[Theorem 2]}. For any $d\otimes d_1\otimes d_2\otimes d_3$ state $\rho_{AB_1B_2B_3}$, we have
\begin{eqnarray}\label{th2}
\mathcal{Q}^{\alpha}_{A|B_1B_2B_3}\geq \sum_{i=1}^{3}\mathcal{Q}^{\alpha}_{AB_i}+\mathcal{Q}^{\alpha}_{A|B^\prime_1|B^\prime_2},
\end{eqnarray}
for $\alpha\geq\beta$.

{\sf[Proof]}. By definition we have
\begin{eqnarray*}\label{}
 \sum_{i=1}^{3}\mathcal{Q}^{\alpha}_{AB_i}+\mathcal{Q}^{\alpha}_{A|B^\prime_{1}|B^\prime_{2}}
&&=\mathcal{Q}^{\alpha}_{AB^\prime_3}+\mathcal{Q}^{\alpha}_{A|B^\prime_1B^\prime_2}\\
&&\leq \mathcal{Q}^{\alpha}_{A|B_1B_2B_3},
\end{eqnarray*}
where $B^\prime_3$ is the complementary of $B^\prime_{1}B^\prime_{2}$ in the subsystem $B_1B_2B_3$, the equality is due to the definition of the residual quantum correlation. From (\ref{th1}), we get the inequality.
\hfill \rule{1ex}{1ex}

Since the last term $\mathcal{Q}^{\alpha}_{A|B^\prime_1|B^\prime_2}$ in (\ref{th2}) is semi-positive,
The inequality (\ref{th2}) is always tighter than (\ref{th1}) for such states $\rho_{AB_1B_2B_3}$. Let us consider the following example based on the quantum correlation measure concurrence. First, we give the definition of concurrence. For a bipartite pure state $|\psi\rangle_{AB}\in H_A\otimes H_B$, the concurrence is $C(|\psi\rangle_{AB})=\sqrt{{2\left[1-\mathrm{Tr}(\rho_A^2)\right]}}$,
where $\rho_A=\mathrm{Tr}_B(|\psi\rangle_{AB}\langle\psi|)$. The concurrence for a bipartite mixed state $\rho_{AB}$ is defined by the convex roof extension,
$C(\rho_{AB})=\min_{\{p_i,|\psi_i\rangle\}}\sum_ip_iC(|\psi_i\rangle)$,
where the minimum is taken over all possible decompositions of $\rho_{AB}=\sum\limits_{i}p_i|\psi_i\rangle\langle\psi_i|$, with $p_i\geq0$ and $\sum\limits_{i}p_i=1$ and $|\psi_i\rangle\in H_A\otimes H_B$. In \cite{ZXN}, the authors show that \begin{eqnarray}\label{zxn}
C^{\alpha}(\rho_{A|B_1B_2\cdots B_{N-1}})\geq C^{\alpha}(\rho_{AB_1})+C^{\alpha}(\rho_{AB_2})+\cdots+C^{\alpha}(\rho_{AB_{N-1}}),
\end{eqnarray} 
for an $N$-qubit state $\rho_{AB_1\cdots B_{N-1}}$.

{\it Example 1}. For the concurrence of the $W$ state,
\begin{eqnarray}\label{W}
|W\rangle_{A|B_1B_2B_3}=\frac{1}{2}(|1000\rangle+|0100\rangle+|0010\rangle+|0001\rangle),
\end{eqnarray}
we have $\beta=2$, $C_{AB_i}=\frac{1}{2}$, $i=1,2,3$, and $C_{A|B_1B_2}=C_{A|B_1B_3}=C_{A|B_2B_3}=\frac{\sqrt{2}}{2}$. Therefore $C^\alpha_{A|B_1|B_2}=C^\alpha_{A|B_1|B_3}=C^\alpha_{A|B_2|B_3}=(\frac{\sqrt{2}}{2})^\alpha-2(\frac{1}{2})^\alpha$. Set $y_1=C^\alpha_{A|B_1B_2B_3}=(\frac{\sqrt{3}}{2})^\alpha,~y_2=\sum_{i=1}^3C^\alpha_{AB_i}=3(\frac{1}{2})^\alpha,~y_3=\sum_{i=1}^3C^\alpha_{AB_i}+C^\alpha_{A|B_1|B_2}=(\frac{\sqrt{2}}{2})^\alpha+(\frac{1}{2})^\alpha$, one can see that our result is better than (\ref{zxn}) in \cite{ZXN}, see Fig. 1.
\begin{figure}
  \centering
  \includegraphics[width=10cm]{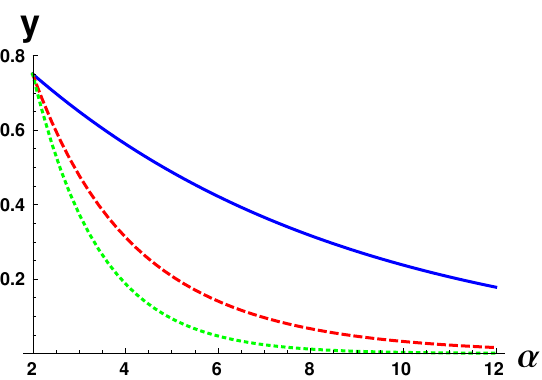}\\
  \caption{Solid (blue) line $y_1$ is the $\alpha$th power of concurrence under bipartition $A|B_1B_2B_3$; Dashed (red) line $y_3$ for the lower bound in (\ref{th2}); Dotted (green) line $y_2$ for the result (\ref{zxn}) in \cite{ZXN}.}\label{2}
\end{figure}

Generalizing the conclusion in Theorem 2 to $N$ partite case, we have the following result.

{\bf[Theorem 3]}. For any $d\otimes d_1\otimes\cdots \otimes d_{N-1}$ state $\rho_{A|B_1B_2\cdots B_{N-1}}$, we have
\begin{eqnarray}\label{th3}
\mathcal{Q}^{\alpha}_{A|B_1B_2\cdots B_{N-1}} \geq \sum_{i=1}^{N-1}\mathcal{Q}^{\alpha}_{AB_i}+\sum_{k=2}^{N-2}\mathcal{Q}^{\alpha}_{A|B^\prime_1|B^\prime_2|\cdots|B^\prime_k},
\end{eqnarray}
for $\alpha\geq\beta$, where $\mathcal{Q}^{\alpha}_{A|B^\prime_1|B^\prime_2|\cdots|B^\prime_{k}}=\mathrm{max}_{1\leq l\leq k+1}\{\mathcal{Q}^{\alpha}_{A|B_{1}|\cdots|\hat{B}_{l}|\cdots|B_{k+1}}\}$ (where $\hat{B}_{l}$ stands for ${B}_{l}$ being omitted in the sub-indices),
$\mathcal{Q}^{\alpha}_{A|B_{1}|B_{2}|\cdots|B_{k+1}}=\mathcal{Q}^{\alpha}_{A|B_1B_2 \cdots B_{k+1}} -\sum_{i=1}^{k+1}\mathcal{Q}^{\alpha}_{AB_i}-\sum_{i=2}^{k}\mathcal{Q}^{\alpha}_{A|B^\prime_1|B^\prime_2|\cdots|B^\prime_i}$,  $2\leq k\leq N-2$, $1\leq l\leq k+1$, $N\geq4$.

{\sf[Proof]}. We prove the theorem by induction. For $N=4$ it reduces to Theorem 2.
Suppose the Theorem 2 holds for $N=n$, i.e.,
\begin{eqnarray}\label{pfth31}
 \mathcal{Q}^{\alpha}_{A|B_1B_2\cdots B_{n-1}} \geq \sum_{i=1}^{n-1}\mathcal{Q}^{\alpha}_{AB_i}+\mathcal{Q}^{\alpha}_{A|B^\prime_{1}|B^\prime_{2}}+\cdots+\mathcal{Q}^{\alpha}_{A|B^\prime_{1}|B^\prime_{2}|\cdots|B^\prime_{n-2}}.
\end{eqnarray}
Then for $N=n+1$, we have
\begin{eqnarray*}\label{pfth32}
&& \sum_{i=1}^{n}\mathcal{Q}^{\alpha}_{AB_i}+\mathcal{Q}^{\alpha}_{A|B^\prime_{1}|B^\prime_{2}}+\cdots+\mathcal{Q}^{\alpha}_{A|B^\prime_{1}|B^\prime_{2}|\cdots|B^\prime_{n-1}}\\\nonumber
&&\leq \mathcal{Q}^{\alpha}_{A|B^\prime_{1}B^\prime_{2}\cdots B^\prime_{n-1}}+\mathcal{Q}^{\alpha}_{AB^\prime_{n}}\\\nonumber
&&\leq \mathcal{Q}^{\alpha}_{A|B_1B_2\cdots B_{n}},
\end{eqnarray*}
where $B^\prime_n$ is the complementary of $B^\prime_{1}B^\prime_{2},\cdots,B^\prime_{n-1}$ in the subsystem $B_1B_2,\cdots,B_n$. The first inequality is due to (\ref{pfth31}). By (\ref{th1}) we get the last inequality.
\hfill \rule{1ex}{1ex}

In Theorems 1 and 2 we have take into account the maximum value among $\mathcal{Q}^{\alpha}_{A|B_{1}|\cdots|\hat{B}_{l}|\cdots|B_{k}}$.
If instead of the maximum value, one just considers the mean value of $\mathcal{Q}^{\alpha}_{A|B_{1}|\cdots|\hat{B}_{l}|\cdots|B_{k}}$,
one may have the following corollary.

{\bf [Corollary 1]}. For any $d\otimes d_1\otimes\cdots \otimes d_{N-1}$ state $\rho_{A|B_1B_2\cdots B_{N-1}}$, we have
\begin{eqnarray}\label{co1}
\mathcal{Q}^{\alpha}_{A|B_1B_2\cdots B_{N-1}}\geq \sum_{i=1}^{N-1}\mathcal{Q}^{\alpha}_{AB_i}+\sum_{k=3}^{N-1}\left(\frac{1}{k}\sum_{l=1}^k\mathcal{Q}^\alpha_{A|B_{1}|\cdots|\hat{B}_{l}|\cdots|B_{k}}\right),
\end{eqnarray}
for all $\alpha\geq\beta$, $N\geq 4$, where
\begin{eqnarray}\label{col1}
 \mathcal{Q}^{\alpha}_{A|B_{1}|B_{2}|\cdots|B_{j}}=\mathcal{Q}^{\alpha}_{A|B_1B_2\cdots B_{j}} -\sum_{i=1}^{j}\mathcal{Q}^{\alpha}_{AB_i}-\sum_{k=3}^{j}\left(\frac{1}{k}\sum_{l=1}^k\mathcal{Q}^\alpha_{A|B_{1}|\cdots|\hat{B}_{l}|\cdots|B_{k}}\right),
\end{eqnarray}
$3\leq j\leq N-1$, $3\leq k\leq N-1$ and $1\leq l\leq k$.

{\it Example 2}. Let us consider the concurrence of the four-qubit pure state,
\begin{eqnarray}\label{FS}
|\psi\rangle_{ABCD}=\frac{1}{\sqrt{3}}(|0000\rangle+|0010\rangle+|1011\rangle).
\end{eqnarray}
We have $\rho_{ACD}=\mathrm{Tr}_{B}(|\psi\rangle_{ABCD}\langle\psi|)=\frac{1}{3}(|000\rangle+|010\rangle+|111\rangle)(\langle000|+\langle010|+\langle111|)$, $\rho_{BCD}=\mathrm{Tr}_{A}(|\psi\rangle_{ABCD}\langle\psi|)=\frac{1}{3}(|000\rangle\langle000|+|000\rangle\langle010|+|010\rangle\langle000|+|010\rangle\langle010|+|011\rangle\langle011|)$, $C_{AB}=C_{AC}=0$, $C_{AD}=\frac{2}{3}$, $C_{BC}=C_{BD}=0$, $C_{A|BC}=0$, $C_{A|BD}=\frac{2}{3}$, $C_{A|CD}=\frac{2\sqrt{2}}{3}$. Therefore, $C_{A|B|C}=C_{A|B|D}=0,~C^\alpha_{A|C|D}=(\frac{2\sqrt{2}}{3})^{\alpha}-(\frac{2}{3})^{\alpha}$. Set $y_1=C^\alpha_{A|BCD}=(\frac{2\sqrt{2}}{3})^{\alpha}$, $y_2=C^\alpha_{AB}+C^\alpha_{AC}+C^\alpha_{AD}=(\frac{2}{3})^{\alpha}$, $y_3=C^\alpha_{AB}+C^\alpha_{AC}+C^\alpha_{AD}+\frac{1}{3}\left(C^\alpha_{A|B|C}+C^\alpha_{A|B|D}+C^\alpha_{A|C|D}\right)=(\frac{2}{3})^{\alpha+1}+\frac{1}{3}(\frac{2\sqrt{2}}{3})^{\alpha}$, one can see that our result is better than that in \cite{ZXN}, see Fig. 2.
\begin{figure}
  \centering
  \includegraphics[width=10cm]{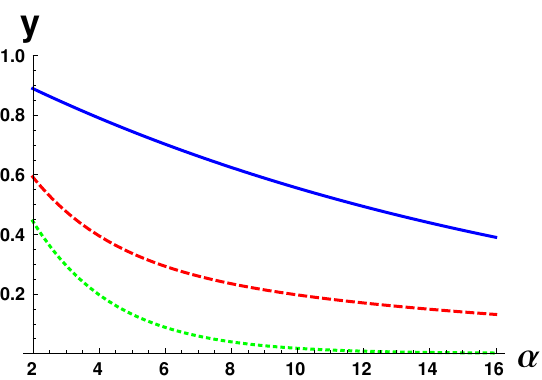}\\
  \caption{Solid (blue) line $y_1$ for the $\alpha$th power of concurrence under bipartition $A|B_1B_2B_3$; Dashed (red) line $y_3$ for the lower bound in (\ref{co1}); Dotted (green) line $y_2$ for the result in \cite{ZXN}. }\label{2}
\end{figure}

Next, we adopt an approach used in \cite{JF} to improve the further above results on monogamy relations for multipartite quantum correlation measures.
First, we give a Lemma.

{\bf [Lemma]}. For any $d_1\otimes d_2\otimes d_3$ mixed state $\rho\in H_A\otimes H_{B}\otimes H_{C}$, if $\mathcal{Q}_{AB}\geq \mathcal{Q}_{AC}$, we have
\begin{equation}\label{la}
  \mathcal{Q}^\alpha_{A|BC}\geq  \mathcal{Q}^\alpha_{AB}+\frac{\alpha}{\beta}\mathcal{Q}^\alpha_{AC},
\end{equation}
for all $\alpha\geq\beta$.

{\sf[Proof]}. For arbitrary $d_1\otimes d_2\otimes d_3$ tripartite state $\rho_{ABC}$.
If $\mathcal{Q}_{AB}\geq \mathcal{Q}_{AC}$, we have
\begin{eqnarray*}
  \mathcal{Q}^\alpha_{A|BC}&&= (\mathcal{Q}^\beta_{AB}+\mathcal{Q}^\beta_{AC})^{\frac{\alpha}{\beta}}=\mathcal{Q}^\alpha_{AB}\left(1+\frac{\mathcal{Q}^\beta_{AC}}{\mathcal{Q}^\beta_{AB}}\right)^{\frac{\alpha}{\beta}} \\
   && \geq \mathcal{Q}^\alpha_{AB}\left[1+\frac{\alpha}{\beta}\left(\frac{\mathcal{Q}^\beta_{AC}}{\mathcal{Q}^\beta_{AB}}\right)^{\frac{\alpha}{\beta}}\right]=\mathcal{Q}^\alpha_{AB}+\frac{\alpha}{\beta}\mathcal{Q}^\alpha_{AC},
\end{eqnarray*}
where the first equality is due to (\ref{aq}), the inequality is due to the inequality $(1+t)^x\geq 1+xt \geq 1+xt^x$ for $x\geq1,~0\leq t\leq1$.
\hfill \rule{1ex}{1ex}

In the above Lemma, without loss of generality, we have assumed that $\mathcal{Q}_{AB}\geq \mathcal{Q}_{AC}$, as the subsystems
$A$ and $B$ are equivalent. Moreover, in the proof of the Lemma we have assumed $\mathcal{Q}_{AB}>0$.
If $\mathcal{Q}_{AB}=0$ and $\mathcal{Q}_{AB}\geq \mathcal{Q}_{AC}$, then $\mathcal{Q}_{AB}=\mathcal{Q}_{AC}=0$. The lower bound is trivially zero.
Generalizing the Lemma to multipartite quantum systems, we have the following Theorem.

{\bf[Theorem 4]}. For any $d\otimes d_1\otimes\cdots \otimes d_{N-1}$ state $\rho\in H_A\otimes H_{B_1}\otimes\cdots\otimes H_{{B_{N-1}}}$, if
${\mathcal{Q}_{AB_i}}\geq {\mathcal{Q}_{A|B_{i+1}\cdots B_{N-1}}}$ for $i=1, 2, \cdots, m$, and
${\mathcal{Q}_{AB_j}}\leq {\mathcal{Q}_{A|B_{j+1}\cdots B_{N-1}}}$ for $j=m+1,\cdots,N-2$,
$\forall$ $1\leq m\leq N-3$, $N\geq 4$, we have
\begin{eqnarray}\label{th4}
\mathcal{Q}^\alpha_{A|B_1B_2\cdots B_{N-1}}\geq &&\mathcal{Q}^\alpha_{AB_1}
+\frac{\alpha}{\beta} \mathcal{Q}^\alpha_{AB_2}+\cdots+\left(\frac{\alpha}{\beta}\right)^{m-1}\mathcal{Q}^\alpha_{AB_m}\\\nonumber
&&+\left(\frac{\alpha}{\beta}\right)^{m+1}(\mathcal{Q}^\alpha_{AB_{m+1}}
 +\cdots+\mathcal{Q}^\alpha_{AB_{N-2}})
+\left(\frac{\alpha}{\beta}\right)^{m}\mathcal{Q}^\alpha_{AB_{N-1}},
\end{eqnarray}
for all $\alpha\geq\beta$.

{\sf[Proof]}. By using the Lemma repeatedly, one gets
\begin{eqnarray}\label{pfth41}
 \mathcal{Q}^{\alpha}_{A|B_1B_2\cdots B_{N-1}}&&\geq \mathcal{Q}^{\alpha}_{AB_1}+\frac{\alpha}{\beta}\mathcal{Q}^{\alpha}_{A|B_2\cdots B_{N-1}}\\\nonumber
&&\geq \mathcal{Q}^{\alpha}_{AB_1}+\frac{\alpha}{\beta}\mathcal{Q}^{\alpha}_{AB_2}
 +\left(\frac{\alpha}{\beta}\right)^2\mathcal{Q}^{\alpha}_{A|B_3\cdots B_{N-1}}\\ \nonumber
 &&\geq\cdots\geq \mathcal{Q}^{\alpha}_{AB_1}+\frac{\alpha}{\beta}\mathcal{Q}^{\alpha}_{AB_2}+\cdots\\\nonumber
 &&+ \left(\frac{\alpha}{\beta}\right)^{m-1}\mathcal{Q}^{\alpha}_{AB_m}
 +\left(\frac{\alpha}{\beta}\right)^m \mathcal{Q}^{\alpha}_{A|B_{m+1}\cdots B_{N-1}}.
\end {eqnarray}
As ${\mathcal{Q}_{AB_j}}\leq {\mathcal{Q}_{A|B_{j+1}\cdots B_{N-1}}}$ for $j=m+1,\cdots,N-2$, by (\ref{pfth41}) we get
\begin{eqnarray}\label{pfth42}
\mathcal{Q}^{\alpha}_{A|B_{m+1}\cdots B_{N-1}}&&\geq \frac{\alpha}{\beta}\mathcal{Q}^{\alpha}_{AB_{m+1}}+\mathcal{Q}^{\alpha}_{A|B_{m+2}\cdots B_{N-1}}\nonumber\\
&&\geq \frac{\alpha}{\beta}(\mathcal{Q}^{\alpha}_{AB_{m+1}}+\cdots+\mathcal{Q}^{\alpha}_{AB_{N-2})}+\mathcal{Q}^{\alpha}_{AB_{N-1}}.
\end{eqnarray}
Combining (\ref{pfth41}) and (\ref{pfth42}), we have Theorem 4.
\hfill \rule{1ex}{1ex}

Similar to the Theorem 3, (\ref{th4}) can be improved by adding a term for residual quantum correlation.
By a similar derivation to Theorem 3, we have

{\bf[Theorem 5]}. For any $d\otimes d_1\otimes\cdots \otimes d_{N-1}$ state $\rho\in H_A\otimes H_{B_1}\otimes\cdots\otimes H_{{B_{N-1}}}$, if
${\mathcal{Q}_{AB_i}}\geq {\mathcal{Q}_{A|B_{i+1}\cdots B_{N-1}}}$ for $i=1, 2, \cdots, m$, and
${\mathcal{Q}_{AB_j}}\leq {\mathcal{Q}_{A|B_{j+1}\cdots B_{N-1}}}$ for $j=m+1,\cdots,N-2$,
$\forall$ $1\leq m\leq N-3$, $N\geq 4$, we have
\begin{eqnarray}\label{}\label{th5}
\mathcal{Q}^\alpha_{A|B_1B_2\cdots B_{N-1}}\geq &&\mathcal{Q}^\alpha_{AB_1}
+\frac{\alpha}{\beta} \mathcal{Q}^\alpha_{AB_2}+\cdots+\left(\frac{\alpha}{\beta}\right)^{m-1}\mathcal{Q}^\alpha_{AB_m}\\\nonumber
&&+\left(\frac{\alpha}{\beta}\right)^{m+1}(\mathcal{Q}^\alpha_{AB_{m+1}}
 +\cdots+\mathcal{Q}^\alpha_{AB_{N-2}})
+\left(\frac{\alpha}{\beta}\right)^{m}\mathcal{Q}^\alpha_{AB_{N-1}}\\\nonumber
&&+\sum_{k=2}^{N-2}\mathcal{\hat{Q}}^{\alpha}_{A|B^\prime_1|B^\prime_2|\cdots|B^\prime_k}\\\nonumber
&&=\sum_{i=1}^{N-1}\mathcal{\hat{Q}}^{\alpha}_{AB_i}+\sum_{k=2}^{N-2}\mathcal{\hat{Q}}^{\alpha}_{A|B^\prime_1|B^\prime_2|\cdots|B^\prime_{k}},
\end{eqnarray}
for all $\alpha\geq\beta$,
where for simplicity, we have denoted $\mathcal{\hat{Q}}^\alpha_{AB_1}=\mathcal{Q}^\alpha_{AB_1}$, $\mathcal{\hat{Q}}^\alpha_{AB_2}=\frac{\alpha}{\beta} \mathcal{Q}^\alpha_{AB_2}$, $\cdots$, $\mathcal{\hat{Q}}^\alpha_{AB_m}=\left(\frac{\alpha}{\beta}\right)^{m-1}\mathcal{Q}^\alpha_{AB_m}$, $\mathcal{\hat{Q}}^\alpha_{AB_{m+1}}=(\frac{\alpha}{\beta})^{m+1}\mathcal{Q}^\alpha_{AB_{m+1}}$, $\cdots$, $\mathcal{\hat{Q}}^\alpha_{AB_{N-2}}=\left(\frac{\alpha}{\beta}\right)^{m+1}\mathcal{Q}^\alpha_{AB_{N-2}}$, $\mathcal{\hat{Q}}^\alpha_{AB_{N-1}}=\left(\frac{\alpha}{\beta}\right)^{m}\mathcal{Q}^\alpha_{AB_{N-1}}$.
The residual quantum correlation term $\mathcal{\hat{Q}}^{\alpha}_{A|B^\prime_1|B^\prime_2|\cdots|B^\prime_{k-1}}=\mathrm{max}_{1\leq l\leq k}\{\mathcal{\hat{Q}}_{A|B_{1}|\cdots|\hat{B}_{l}|\cdots|B_{k}}\}$,
$\mathcal{\hat{Q}}^{\alpha}_{A|B_{1}|B_{2}|\cdots|B_{k}}=\mathcal{Q}^{\alpha}_{A|B_1B_2\cdots B_{k}} -\sum_{i=1}^{k}\mathcal{\hat{Q}}^{\alpha}_{AB_i}-\sum_{i=2}^{k-1}\mathcal{\hat{Q}}^{\alpha}_{A|B^\prime_1|B^\prime_2|\cdots|B^\prime_i}$, $2\leq k\leq N-2$, $1\leq l\leq k$.

As an example, let us consider consider again the the concurrence of the state (\ref{W}). We have
$\hat{C}^\alpha_{A|B_1|B_2}=\hat{C}^\alpha_{A|B_1|B_3}=\hat{C}^\alpha_{A|B_2|B_3}=(\frac{\sqrt{2}}{2})^\alpha-(1+\frac{\alpha}{2})(\frac{1}{2})^\alpha$. Set $y_1=C^\alpha_{A|B_1B_2B_3}=(\frac{\sqrt{3}}{2})^\alpha$, $y_2=\sum_{i=1}^3\hat{C}^\alpha_{AB_i}+\hat{C}^\alpha_{A|B_1|B_2}=(\frac{\sqrt{2}}{2})^\alpha+\frac{\alpha}{2}(\frac{1}{2})^\alpha$, $y_3=\sum_{i=1}^3\hat{C}^\alpha_{AB_i}=(\alpha+1)(\frac{1}{2})^\alpha$. We see in Fig. 3 that the bound (\ref{th4}) is improved.

\begin{figure}
  \centering
  \includegraphics[width=10cm]{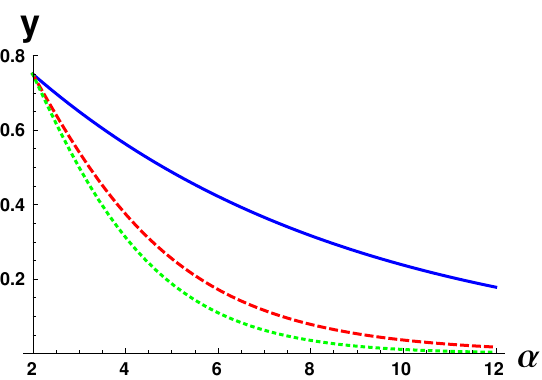}\\
  \caption{Solid (blue) line $y_1$ is the $\alpha$th power of concurrence under bipartition $A|B_1B_2B_3$. The dashed (red) line $y_2$ is the lower bound in (\ref{th5}) and the dotted (green) $y_3$ line for (\ref{th4}).}\label{2}
\end{figure}

In the following, we consider the special multi-qubit, $d=d_1=\cdots=d_{N-1}=2$, for which one may have richer results.
In \cite{GSB}, it has been shown that 
\begin{eqnarray}\label{ca}
C^2_{A|B_1B_2\cdots B_{N-1}}\leq {C^2_a}_{AB_1}+{C^2_a}_{AB_2}+\cdots+{C^2_a}_{AB_{N-1}},
\end{eqnarray}
where the concurrence of assistance is defined as $C_a(|\psi\rangle_{ABC})\equiv C_a(\rho_{AB})=\max\limits_{\{p_i,|\psi_i\rangle\}}\sum_ip_iC(|\psi_i\rangle)$, with the maximum is taken over all possible decompositions of $\rho_{AB}=\mathrm{Tr}_C(|\psi\rangle_{ABC}\langle\psi|)=\sum\limits_{i}p_i|\psi_i\rangle_{AB}\langle\psi_i|$, denote ${C_a}_{AB}=C_a(\rho_{AB})$.

The residual quantum correlation for the concurrence can be also used to improve other kinds of
monogamous relations based on concurrence and concurrence of assistance \cite{ca}.
For $N$-qubit systems $ABC_1\cdots C_{N-2}$, the monogamy relations satisfied by the concurrence of $N$-qubit pure states under the partition $AB$ and $C_1 . . . C_{N-2}$
have been first time established in \cite{ZXN2}. In following we give an improved one.

{\bf [Theorem 6]}. For any $2\otimes2\otimes\cdots\otimes2$ pure state $|\psi\rangle_{ABC_1\cdots C_{N-2}}$, denote $\rho_{AC_0}=\rho_{AB},~\rho_{BC_0}=\rho_{BA}$, we have
\begin{eqnarray}\label{th7}
  C^2_{AB|C_1\cdots C_{N-2}}\geq\mathrm{max}
\left\{
   \begin{aligned}
   \sum_{i=0}^{N-2}(C^2_{AC_i}-{C^2_a}_{BC_i})+\sum_{k=1}^{N-3}C^2_{A|C^\prime_0|C^\prime_1|\cdots|C^\prime_{k}}, \\
   \sum_{i=0}^{N-2}(C^2_{BC_i}-{C^2_a}_{AC_i})+\sum_{k=1}^{N-3}C^2_{B|C^\prime_0|C^\prime_1|\cdots|C^\prime_{k}},\\
   \end{aligned}
   \right.
  \end{eqnarray}
where $C^2_{A|C^\prime_0|C^\prime_1|\cdots|C^\prime_{k}}=\mathrm{max}_{0\leq l\leq k+1}\{C^2_{A|C_{0}|\cdots|\hat{C}_{l}|\cdots|C_{k+1}}\}$,
$C^2_{A|C_{0}|C_{1}|\cdots|C_{k}}=C^2_{A|C_0C_1\cdots C_{k}} -\sum_{i=0}^{k}C^2_{AC_i}-\sum_{i=2}^{k-1}C^2_{A|C^\prime_0|C^\prime_1|\cdots|C^\prime_i}$, $1\leq k\leq N-3$, $0\leq l\leq k+1$.

{\sf [Proof]}. For $2\otimes2\otimes\cdots\otimes2$ state $|\psi\rangle_{ABC_1\cdots C_{N-2}}$, one has
\begin{eqnarray*}
C^2_{AB|C_1\cdots C_{N-2}}&&=2T(\rho_{AB})\\
&&\geq2T(\rho_A)-2T(\rho_{B})\\
&&=C^2_{A|BC_1\cdots C_{N-2}}-C^2_{B|AC_1\cdots C_{N-2}}\\
&&\geq \sum_{i=0}^{N-2}C^2_{AC_i}+\sum_{k=1}^{N-3}C^2_{A|C^\prime_0|C^\prime_1|\cdots|C^\prime_k}-C^2_{B|AC_1\cdots C_{N-2}}\\
&&\geq \sum_{i=0}^{N-2}C^2_{AC_i}+\sum_{k=1}^{N-3}C^2_{A|C^\prime_0|C^\prime_1|\cdots|C^\prime_k}- \sum_{i=0}^{N-2}{C^2_a}_{BC_i},
\end{eqnarray*}
where $T(\rho)=1-\mathrm{Tr}(\rho^2)$, and the first inequality is due to a property of the linear entropy. Using the Theorem 3, one can get the second inequality. The last inequality is obtained by (\ref{ca}).
\hfill \rule{1ex}{1ex}

The last term in (\ref{th7}) improves the result in \cite{ZXN2}. Consider the concurrence of (\ref{FS}), $|\psi\rangle_{AB_1B_2B_3}=|\psi\rangle_{ABCD}$. By the Theorem 6,
we have $C_{AB|CD}\geq \frac{8}{9}$, which is better than the result $C_{AB|CD}\geq \frac{4}{9}$ in \cite{ZXN2}.

Now we generalize our results to the concurrence $C_{ABC_1|C_2\cdots C_{N-2}}$ under partition $ABC_1$ and $C_2 \cdots C_{N-2}~(N\geq6)$ for pure state $|\psi\rangle_{ABC_1\cdots C_{N-2}}$. Similar to Theorem 6, we can obtain the following corollary:

{\bf [Corollary 2]}.  For any $N$-qubit pure state $|\psi\rangle_{ABC_1\cdots C_{N-2}}$, we have
\begin{eqnarray}\label{co2}
C^2_{ABC_1|C_2\cdots C_{N-2}}\geq\mathrm{max}\left\{J_A, J_B\right\}-J_{C_1},
\end{eqnarray}
where $J_A=\sum_{i=0}^{N-2}(C^2_{AC_i}-{C^2_a}_{BC_i})+\sum_{k=1}^{N-3}C^2_{A|C^\prime_0|C^\prime_1|\cdots|C^\prime_{k}}$, $J_B=\sum_{i=0}^{N-2}(C^2_{BC_i}-{C^2_a}_{AC_i})+\sum_{k=1}^{N-3}C^2_{B|C^\prime_0|C^\prime_1|\cdots|C^\prime_{k}}$, $J_{C_1}={C^2_a}_{C_1A}+{C^2_a}_{C_1B}+\sum_{i=2}^{N-2}{C^2_a}_{C_1C_i}$.

{\sf [Proof]}. For any $N$-qubit pure state $|\psi\rangle_{ABC_1\cdots C_{N-2}}$, we have
\begin{eqnarray*}
C^2_{ABC_1|C_2\cdots C_{N-2}}&&=2T(\rho_{ABC_1})\\
&&\geq2T(\rho_{AB})-2T(\rho_{C_1})\\
&&= C^2_{AB|C_1\cdots C_{N-2}}-C^2_{C_1|ABC_2\cdots C_{N-2}},
\end{eqnarray*}
where the inequality is due to the property of the linear entropy $T(\rho_{ABC_1})\geq T(\rho_{AB})-T(\rho_{C_1})$. Combining (\ref{ca}) and (\ref{th7}), we obtain (\ref{co2}).
\hfill \rule{1ex}{1ex}

\smallskip

We have presented general monogamy relations for any quantum correlation measures and multipartite quantum states.
Similar to the three tangle of concurrence, we defined the $\alpha$th $(\alpha\geq\beta)$ power of the residual quantum correlation.
Based on this, we have established tighter monogamy inequalities for arbitrary quantum correlation measures.
For qubit systems, the bound for concurrence, given by concurrence of assistance, has been also improved.
Finally, we have presented a different kind of monogamy relations satisfied by the concurrence of $N$-qubit pure states under partition $AB$ and $C_1\cdots C_{N -2}$,
as well as under partition $ABC_1$ and $C_2 \cdots C_{N-2}$, which is also shown to be better than the existing ones.
The residual quantum correlation we introduced may also contribute to improve other relations satisfied by the measures of quantum correlations.


\begin{thebibliography}{99}
 \bibitem{CS} C. H. Bennett, S. J. Wiesner, Phys. Rev. Lett 69 (1992) 2881.
 \bibitem{CGC} C. H. Bennett, G. Brassard, C. Cr$\acute{\mathrm{e}}$peau, R. Josza, A. Peres, W.
 K. Wootters, Phys. Rev. Lett 70 (1993) 1895.
 \bibitem{JZC} J. W. Pan, Z. B. Chen, C. Y. Lu, H. Weinfurter, A. Zeilinger, M. $\dot{\mathrm{Z}}$ukowski, Rev. Mod. Phys. 84 (2012) 777-838.
 \bibitem{FFF} A. Sen(De), U. Sen, Physics News 40 (2010) 17-32 (available at arXiv:quant- ph/1105.2412).
 \bibitem{MAV} M. Lewenstein, A. Sanpera, V. Ahufinger, B. Damski, A. Sen(De), U. Sen, Adv. Phys 56 (2007) 243–379.
 \bibitem{LRA} L. Amico, R. Fazio, A. Osterloh, V. Vedral, Rev. Mod. Phys 80 (2008) 517-576.


\bibitem{rpmk}  R. Horodecki, P. Horodecki, M. Horodecki, K. Horodecki, Rev. Modern Phys. 81 (2009) 865–942.
\bibitem{kah} K. Modi, A. Brodutch, H. Cable, T. Paterek, V. Vedral, Rev. Modern Phys. 84 (2012) 1655.




\bibitem{AKE} A. K. Ekert, Phys. Rev. Lett. 67 (1991) 661; V. Coffman, J. Kundu, W. K. Wootters, Phys. Rev. A 61 (2000) 052306.
\bibitem{CBHSB} C. H. Bennett, H. J. Bernstein, S. Popescu, B. Schumacher, Phys. Rev. A 53 (1996) 2046.
\bibitem{KKA} M. Koashi, A. Winter, Phys. Rev. A 69 (2004) 022309.
\bibitem{SSS} G. Adesso, A. Serafini, F. Illuminati, Phys. Rev. A 73 (2006) 032345.

\bibitem{bt} B. Toner, Proc. R. Soc. A 465, 59 (2009). 

\bibitem{bf} B. Toner and F, arXiv:quant-ph/0611001.

\bibitem{mdr} M. D. Reid, Phys. Rev. A 88, 062108 (2013). 

\bibitem{hqy} Y. Xiang, I. Kogias, G. Adesso, and Q. Y. He, Phys. Rev. A 95, 010101 (2017).

\bibitem{csm} S. Cheng, A. Milne, M. J. W. Hall, and H. M. Wiseman, Phys. Rev. A 94, 042105 (2016). 

\bibitem{am} A. Milne, S. Jevtic, D. Jennings, H. Wiseman, and T. Rudolph, New J. Phys. 16, 083017 (2014). 








\bibitem{NGWH} N. Gisin, G. Ribordy, W. Tittel, H. Zbinden, Rev. Mod. Phys 74 (2002) 145-195;\\
B. M. Terhal, IBM J. Res. Dev. 48 (2004) 71-78.
\bibitem{concurrence}
S. Hill, W. K. Wootters, Phys. Rev. Lett 78 (1997) 5022-5025.\\
S. Albeverio, S. M. Fei. J Opt B: Quantum Semiclass Opt 3 (2001) 223-227.

\bibitem{ZXN} X. N. Zhu and S. M. Fei, Phys. Rev. A 90 (2014) 024304.
\bibitem{JF} Z. X. Jin and S. M. Fei, Quant. Inf. Process (2017) 16:77.
\bibitem{HOWH} H. Ollivier, W. H. Zurek, Phys. Rev. Lett 88 (2001) 017901.
\bibitem{RPAK} R. Prabhu, A. K. Pati, A. Sen(De), U. Sen, Phys. Rev. A 85 (2012) 040102(R).
\bibitem{GLGP} G. L. Giorgi, Phys. Rev. A 84 (2011) 054301.
\bibitem{RPAKA} R. Prabhu, A. K. Pati, A. Sen(De), U. Sen, Phys. Rev. A 86 (2012) 052337.
\bibitem{XJRHF} X. J. Ren, H. Fan, Quant. Inf. Comp 13 (2013) 0469-0478.
\bibitem{YKBN} Y. K. Bai, N. Zhang, M. Y. Ye, Z. D. Wang, Phys. Rev. A 88 (2013) 012123.


\bibitem{ARA} A. Kumar, R. Prabhu, A. Sen(De), and U. Sen, Phys. Rev. A 91 (2015) 012341.

\bibitem{SPAU}K. Salini, R. Prabhu, A. Sen(De), and U. Sen. Ann. Phys 348 (2014) 297-305. 
\bibitem{ca} G. Gour, D. A. Meyer, and B. C. Sanders, Phys. Rev. A 72 (2005) 042329. 
\bibitem{GSB} G. Gour, S. Bandyopadhyay, and B. C. Sanders, J. Math. Phys 48 (2007) 012108.
\bibitem{ZXN2} X. N. Zhu and S. M. Fei, Phys. Rev. A 92 (2015) 062345.
\end{thebibliography}
\end{document}